\documentstyle[preprint,aps,version2]{revtex}
\draft
\def\Journal#1#2#3#4{{#1} {#2} {(#4)} {#3}}

\def\NP{{ Nucl. Phys.} }
\def\PLB{{ Phys. Lett.}  B}
\def\PRL{ Phys. Rev. Lett.}
\def\PRD{{ Phys. Rev.} D}
\def\ZPC{{Z. Phys.} C}
\def\EPJC{{Eur. Phys. J.} C}

\def\CPC{Comput. Phys. Commun.}

\def\ra{\rightarrow}

\def\be{\begin{equation}}
\def\ee{\end{equation}}
\def\bea{\begin{eqnarray}}
\def\eea{\end{eqnarray}}

\def\qbar{{\bar q}}
\def\ubar{{\bar u}}
\def\dbar{{\bar d}}
\def\sbar{{\bar s}}

\def\NP{{ Nucl. Phys.}}
\def\ANP{{Adv. Nucl. Phys.}}
\def\PR{{Phys. Rep}}

\begin{document}
\begin{titlepage} 

\bigskip
\begin{title}
{\large\bf The flavour asymmetry and quark-antiquark asymmetry in the $\Sigma^+$-sea}

\end{title}
\author{Fu-Guang Cao\thanks{E-mail address: f.g.cao@massey.ac.nz.}
 and A. I. Signal\thanks{E-mail address: a.i.signal@massey.ac.nz.}}
\begin{instit}
Institute of Fundamental Sciences, Massey University,
Private Bag 11 222, Palmerston North,
New Zealand
\end{instit}

\begin{abstract}

The sea quark content of the $\Sigma^+$ baryon is investigated using
light-cone baryon-meson fluctuation model suggested
by Brodsky and Ma.
It is found that the $\Sigma^+$ sea is flavour asymmetric
($\dbar > \ubar > \sbar$) and quark-antiquark asymmetric ($q \not= \qbar$).
Our prediction for the flavour asymmetry, $\dbar > \ubar > \sbar$, is significantly
different from the $SU(3)$ prediction ($\dbar < \ubar < \sbar$),
while our prediction for the $d$-$\dbar$ asymmetry is consistent with the
$SU(3)$ prediction.


\end{abstract}
\end{titlepage}

The nucleon sea exhibits two interesting properties: flavour asymmetry
\cite{New,NA51,E866,Hermes} and quark-antiquark
asymmetry \cite{CCFR,NMC}.
While there have been many studies of the nucleon sea from both experiment
(see e.~g. \cite{New,NA51,E866,Hermes,CCFR,NMC}) and
theory (see e.~g. \cite{Thomas83,Holtmann,Speth,Kumano,Boros} and references therein),
the studies of the sea distributions of the other baryons
in the baryon octet predicted by the $SU(3)$ quark model are very few.
It is of interest to know whether the sea of the other members of the
baryon octet has the same properties
(flavour asymmetry and quark-antiquark asymmetry) as the nucleon sea.
Also, through the study of the quark sea of the
other members of the baryon octet,
we can improve our understanding of
the structure of the baryons and the non-perturbative properties of QCD.
Alberg {\it et al.} \cite{Alberg} pointed out that
the valence and sea quark distributions of the $\Sigma^{\pm}$
may exhibit large deviations from the $SU(3)$ predictions,
and these parton distributions
could be obtained from Drell-Yan experiments using charged hyperon
beams on proton and deuteron targets.
Alberg, Falter, and Henley \cite{Alberg2} studied the flavour asymmetry
in the $\Sigma^+$ sea employing the meson cloud model
and effective Lagrangian for the baryon-meson-baryon interaction,
and found large deviations from $SU(3)$.
Boros and Thomas \cite{BorosThomas} calculated the quark distributions of
$\Lambda$ and $\Sigma^{\pm}$ employing the MIT bag model.
It was found that the valence quark distributions are quite different from
the $SU(3)$ predictions and that the quark sea is flavour asymmetric.
More recently, Ma, Schmidt and Yang \cite{MaSY} showed that
there are significant differences between the predictions
of perturbative QCD and $SU(6)$ quark-diquark model for the flavor
and spin structure of the $\Lambda$ baryon's quark distributions near $x=1$.

In this letter we shall investigate the flavour asymmetry and quark-antiquark
asymmetry of the $\Sigma^+$ sea using the light-cone baryon-meson
fluctuation model (LCM) suggested by Brodsky and Ma \cite{BrodskyMa}.
The baryon-meson fluctuation (meson cloud) mechanism is very successful
in understanding on the flavour asymmetry and quark-antiquark
asymmetry of nucleon sea.
The various fluctuations can be described via corresponding
baryon-meson-nucleon Lagrangians \cite{Thomas83,Holtmann,Speth,Kumano}.
Recently, Brodsky and Ma \cite{BrodskyMa} proposed that
the baryon-meson fluctuation could be described
by using a light-cone two body wave function which is a function of
the invariant mass squared of the baryon-meson Fock state.
Compared to the commonly used effective Lagrangian method (ELM)
\cite{Thomas83,Holtmann,Speth,Kumano} for the description
of baryon-meson fluctuations, the LCM is relatively simple.
Furthermore our study \cite{Cao} showed that the LCM can produce
very similar results to the effective Lagrangian method
for a suitable choice of parameter.

The basic idea of the meson cloud model
(for recent reviews see Refs.~\cite{Speth,Kumano}) is that the nucleon
can be viewed as a bare nucleon surrounded by a mesonic cloud.
The nucleon wave function can be expressed in terms of bare nucleon and
virtual baryon-meson Fock states.
Although this model was developed mainly in the study of nucleon sea,
applying this model to the other baryons is straightforward.
For the $\Sigma^+$, the wave function can be written as
\bea
|\Sigma^+\rangle_{\rm physical} = Z |\Sigma^+\rangle_{\rm bare}
+ \sum_{B M} \int dy \, d^2 {\bf k}_\perp \, \phi_{B M}(y,k_\perp^2)
       \, |{B}(y, {\bf k}_\perp); {M}(1-y,-{\bf k}_\perp)\rangle,
\label{NMCM}
\eea
where $Z$ is the wave function renormalization constant,
$\phi_{BM}(y,k_\perp^2)$ is the wave function
of Fock state containing a baryon ($B=\Lambda$, $\Sigma^0$, $\Sigma^+$, $p$)
with longitudinal momentum fraction $y$, transverse momentum ${\bf k}_\perp$,
and a meson ($M=\pi^+$, $\pi^0$, ${\bar K^0}$) with momentum fraction $1-y$,
transverse momentum $-{\bf k}_\perp$.
Here we consider the most energetically-favoured fluctuations
in the baryon octet and meson octet.
The fluctuation $\Sigma^+ \ra \Xi^0 K^+$ is neglected due to the
higher mass of $\Xi^0$ ($m_{\Xi}=1.32$ GeV while $m_\Lambda=1.12$ GeV,
$m_\Sigma=1.19$ GeV).

It would seem that the fluctuation $\Sigma^+ \ra \Sigma^+ \eta$
is also important in the calculations of $\dbar - \sbar$ and $\ubar - \sbar$.
However, applying the common $\eta_8$-$\eta_1$ mixing scheme
\bea
\eta={\rm cos} \, \theta \frac{1}{\sqrt{6}}(u \ubar + d \dbar -2 s \sbar)
       -{\rm sin} \,  \theta \frac{1}{\sqrt{3}}(u \ubar + d \dbar + s \sbar)
\eea
and assuming $SU(3)$ symmetry for the quark distributions in
the $\eta_8$ and $\eta_1$,
we find that compared to the fluctuation $\Sigma^+ \ra \Lambda \pi^+$,
the contributions to the $\dbar -\sbar$ and $\ubar-\sbar$
from the fluctuation $\Sigma^+ \ra \Sigma^+ \eta$
are suppressed by a factor of
$(\frac{1}{\sqrt{6}}{\rm cos} \theta -\frac{1}{\sqrt{3}} {\rm sin} \theta)^2
-(\frac{2}{\sqrt{6}}{\rm cos} \theta +\frac{1}{\sqrt{3}} {\rm sin} \theta)^2$
which is in the range of $-0.20 \sim -0.01$ for mixing angle in
the theoretically accepted range $\theta=-12^o \sim -20^o$
\cite{Schechter,Feldmann,CaoMixing,Donoghue,Burakovsky}.
The higher mass of the $\eta$ ($m_\eta=0.547$ GeV, $m_\pi=0.139$ GeV)
also suppresses the contribution from this fluctuation.
Thus we neglect this fluctuation in our calculation.

Provided that the lifetime of a virtual baryon-meson Fock state is much longer
than the strong interaction time in the Drell-Yan process,
the contribution from the virtual baryon-meson Fock states to the quark and
anti-quark sea of $\Sigma^+$ can be written as convolutions
\bea
q(x)&=&\sum_{BM} \left[\int^1_x \frac{dy}{y} f_{BM}(y) q^B(\frac{x}{y})
		   +\int^1_x \frac{dy}{y} f_{MB}(1-y) q^M(\frac{x}{y}) \right],
\label{qBM} \\
\qbar(x)&=&\sum_{BM} \int^1_x \frac{dy}{y} f_{MB}(1-y) \qbar^M(\frac{x}{y}),
\label{qbarBM}
\eea
where $f_{B M}(y)=f_{MB}(1-y)$ is fluctuation function which gives
the probability for the $\Sigma^+$ to fluctuate into a virtual $BM$ state
\bea
f_{BM}(y)=\int^\infty_0 d k_\perp^2 \left | \phi_{B M}(y, k_\perp^2)\right |^2.
\label{fBM}
\eea

A common practice in the evaluation of the wave function
$\phi_{B M}(y, {\bf k}^2_\perp)$ is to employ time-ordered perturbative theory
in the infinite momentum frame and the effective meson-baryon-nucleon
interaction Lagrangian \cite{Thomas83,Holtmann,Speth,Kumano}.
On the other hand, Brodsky and Ma \cite{BrodskyMa} suggested
that this wave function can also be described by using light-cone
two-body wave function which is a function of the
the invariant mass squared of the baryon-meson Fock state
\bea
\phi_{BM}(y,{\bf k}_\perp)=A\, {\rm exp}
\left[\frac{1}{8\alpha^2}\left(\frac{m_B^2+{\bf k}_\perp^2}{y}
+\frac{m_M^2+{\bf k}_\perp^2}{1-y} \right)\right],
\label{phi}
\eea
where $\alpha$ is a phenomenological parameter which determines the shape of
the fluctuation function.
Compared to the effective Lagrangian method,
Eq.~(\ref{phi}) is quite simple.
Furthermore, our study on the $s$-$\sbar$ asymmetry in the nucleon sea \cite{Cao}
showed that Eq.~(\ref{phi}) can provide similar results to
the effective Lagrangian method for $\alpha=1.0$ GeV.
Because the spin structure of the baryon-meson-baryon vertex is the same for
all members of the baryon octet (ignoring fluctuations to decuplet baryons),
we might expect that the value of $\alpha$ should be similar for all the members
of the baryon octet.
We will use $\alpha=0.3$ GeV and $1.0$ GeV in our calculation
as there is little constraint from experimental data or theoretical studies
on the $\Sigma$ sea.
The normalization $A$ in Eq.~(\ref{phi}) can be determined by the probability
for the corresponding fluctuation.
We adopt the result given in Ref.~ \cite{BorosThomas} for the probabilities
of the various fluctuations\footnote{Note the relationship between the
fluctuation functions for various iospin states:
$f_{\Sigma^0 \pi^+}=f_{\Sigma^+\pi^0}$ and the fluctuation functions
given in Ref.~\cite{BorosThomas} for a given type of fluctuation are defined as
the sum of all iospin states:
$f_{\Sigma \pi}=f_{\Sigma^0\pi^+}+f_{\Sigma^+ \pi^0}$.}:
\bea
P_{\Lambda \pi^+}=3.2\%,  \,\,\,\, P_{p {\bar K^0}}=0.4\%, \\
P_{\Sigma^0 \pi^+}=P_{\Sigma^+ \pi^0}=\frac{1}{2}P_{\Sigma \pi}=1.85\%.
\label{P}
\eea

In the baryon-meson fluctuation model, the non-perturbative contributions
to the quark and the anti-quark distributions
in the $\Sigma^+$ sea come from the quarks and anti-quarks of the baryons
($\Lambda, \, \Sigma^+,\, \Sigma^0$ and $p$) and mesons
($\pi^+, \, \pi^0$ and ${\bar K^0}$) in the virtual baryon-meson Fock states.
So we need the parton distributions of the involved baryons and mesons as input.
For the parton distribution in the pion, we employ the parameterization
given by Gl\"{u}ck, Reya, and Stratmann (GRS98) \cite{GRS98}
and we neglect the sea content in the meson, that is, 
\bea
\dbar^{\pi^+}=u^{\pi^+}&=&\ubar^{\pi^-}=d^{\pi^-}=\frac{1}{2} v^\pi, \\
\ubar^{\pi^0}=u^{\pi^0}&=&\dbar^{\pi^0}=d^{\pi^0}=\frac{1}{4} v^\pi, \\
v^\pi(x,\mu_{\rm NLO}^2) &=&1.052 x^{-0.495} (1 +0.357 \sqrt{x}) (1-x)^{0.365},
\label{vpion}
\eea
at scale $\mu_{\rm NLO}^2=0.34$ GeV$^2$.
For the $\dbar$ distribution in the ${\bar K^0}$ we relate it
to the $u$ distribution in the $K^+$
which are given in the GRS98 parameterization \cite{GRS98} also
\bea
\dbar^{{\bar K^0}}(x,\mu_{\rm NLO}^2)=u^{K^+}(x,\mu_{\rm NLO}^2)
= 0.540 (1-x)^{0.17} v^\pi(x,\mu_{\rm NLO}^2),
\label{vK0bar}
\eea
at scale $\mu_{\rm NLO}^2=0.34$ GeV$^2$.

In order to investigate the quark-antiquark asymmetry via $d(x)-\dbar(x)$
in the $\Sigma^+$ sea, we also need to know the $d$-quark distribution in
the $\Lambda$, $\Sigma^+$ and $p$, for which
we use the parameterization for the $d$ quark distribution in the proton
given by Gl\"{u}ck, Reya, and Vogt (GRV98) \cite{GRV98},
\bea
d^p(x,\mu_{\rm NLO}^2)=0.400 \, x^{-0.57} (1-x)^{4.09} (1+18.2 x),
\eea
at scale $\mu_{\rm NLO}^2=0.40$ GeV$^2$.

We evolve the distributions to the scale $Q^2 = 4$ GeV$^2$
using the program of Miyama and Kumano \cite{MiyamaK}
in which the evolution equation is solved numerically in a brute-force method.
We found that at $Q^2=4$ GeV$^2$ all parton distributions
$v^\pi(x, Q^2)$, $\dbar^{{\bar K^0}}(x, Q^2)$ and $d^p(x, Q^2)$
can be parameterized using the following form
\bea
q(x, Q^2)=a \, x^b \, (1-x)^c\, (1+d \, \sqrt{x} +e \, x)
\label{qfit}
\eea
with the parameters given in Table 1.
We estimate the uncertainty in solving the evolution equations numerically
and parameterizating the parton distribution in the form of Eq.~(\ref{qfit})
to be about $2\%$ in the $x$-region which we are interested in
{\it i.e.} $x >10^{-3}$.
The effect of evolution from a lower scale to a higher
scale is to make the parton distribution more concentrated
in the small $x$ region. Thus we may expect that the $x$ position at which
an asymmetry exhibits a maximum will move to
smaller $x$ as we evolve to higher values of $Q^2$.
However, we do not expect the asymmetry
to ``evolve away" at a higher $Q^2$ scale if it exists
at a lower scale such as $\mu^2_{\rm NLO}$.

We investigate the flavour asymmetry in the $\Sigma^+$ sea
through calculating the differences between the antiquark distributions:
$x[\dbar(x)-\ubar(x)]$, $x[\dbar(x)-\sbar(x)]$ and  $x[\ubar(x)-\sbar(x)]$
which are given by
\bea
x\left[\dbar(x)-\ubar(x)\right] &=& x \left[ \dbar_{\Lambda \pi^+}(x)
				        +\dbar_{\Sigma^0 \pi^+}(x)
				        +\dbar_{p {\bar K^0}}(x) \right] \nonumber \\
&=&\int^1_x dy \, \frac{x}{y} \left[
 f_{\Lambda \pi^+}(1-y) \dbar^{\pi^+}(\frac{x}{y})
+f_{\Sigma^0 \pi^+}(1-y) \dbar^{\pi^+}(\frac{x}{y}) \right.\nonumber \\
& &~~~~~~~~~ \left. + f_{p {\bar K^0}}(1-y) \dbar^{\bar K^0}(\frac{x}{y})
\right], 
\label{xdmu}
\eea
\bea
x\left[\dbar(x)-\sbar(x)\right]&=& x \left[\dbar_{\Lambda \pi^+}(x)
				        +\dbar_{\Sigma^0 \pi^+}(x)
				        +\dbar_{\Sigma^+ \pi^0}(x) 
				        +\dbar_{p {\bar K^0}}(x) \right]\nonumber \\
&=&\int^1_x dy \, \frac{x}{y} \left[
 f_{\Lambda \pi^+}(1-y) \dbar^{\pi^+}(\frac{x}{y})
+f_{\Sigma^0 \pi^+}(1-y) \dbar^{\pi^+}(\frac{x}{y}) \right.\nonumber \\
& &~~~~~~~~~~\left. +f_{\Sigma^+ \pi^0}(1-y) \dbar^{\pi^0}(\frac{x}{y})
      + f_{p {\bar K^0}}(1-y) \dbar^{\bar K^0}(\frac{x}{y})
\right],
\label{xdms}
\eea
\bea
x\left[\ubar(x)-\sbar(x)\right]&=& x \, \dbar_{\Sigma^+ \pi^0}(x) \nonumber \\
&=&\int^1_x dy \, \frac{x}{y} f_{\Sigma^+ \pi^0}(1-y) 
     \dbar^{\pi^0}(\frac{x}{y}).
 \label{xums}
\eea
$x [\ubar(x)-\sbar(x)]$ comes from
only fluctuation $\Sigma^+ \ra \Sigma^+ \pi^o$,
while $x [\dbar(x)-\ubar(x)]$ and $x [\dbar(x)-\sbar(x)]$
come from also $\Sigma^+ \ra \Lambda \pi^+$,
$\Sigma^+ \ra \Sigma^0 \pi^+$ as well as $\Sigma^+ \ra p {\bar K^0}$.
In Fig.~1 we present our results for $x [\dbar(x)-\ubar(x)]$
at the scales $\mu^2_{\rm NLO}=0.34$ GeV$^2$ and $Q^2=4$ GeV$^2$
with $\alpha=0.3$ GeV.
It can be found that the contribution from the fluctuation
$\Sigma^+ \ra \Lambda \pi^+$ is about twice as large as that
from $\Sigma^+ \ra \Sigma^0 \pi^+$, and both are much larger
than the contribution from $\Sigma^+ \ra p {\bar K^0}$.
Under evolution the distributions move
from larger $x$ to smaller $x$ --
the $x$ position at which $x[\dbar(x)-\ubar(x)]$ exhibits a maximum shifts
from about $0.1$ to $0.06$
and the maximum decreases about $20\%$,
which coincides with our naive expection.
The numerical results for $x[\dbar(x)-\ubar(x)]$, $x[\dbar(x)-\sbar(x)]$
and $x[\ubar(x)-\sbar(x)]$ at $Q^2=4$ GeV$^2$
are given in Figs.~2 and 3  for $\alpha=0.3$ GeV and $1.0$ GeV respectively.
We can see that $\dbar(x) > \ubar(x) > \sbar(x)$, that is
the anti-quark distribution in the $\Sigma^+$ sea is flavour asymmetric.

As is well known, the nucleon sea is also asymmetric
and for the proton sea $\dbar>\ubar>\sbar$
\cite{Thomas83,Holtmann,Speth,Kumano,Boros}.
The main difference between the proton$(uud)$ and
$\Sigma^+(uus)$ is the replacement of a valance $d$ quark
by a valance $s$ quark.
Thus one may expect from $SU(3)$ symmetry that
the $\Sigma^+(uus)$ sea to be $\sbar>\ubar>\dbar$.
This prediction is opposite to our above conclusion ($\dbar>\ubar>\sbar$)
from the light-cone baryon-meson fluctuation model.
If the $SU(3)$ symmetry breaking in the $\ubar$, $\dbar$ and $\sbar$
distributions in the $\Sigma^+$ sea has the same source as that for
the $u$, $d$, and $s$ quark masses,
we may expect that $x[\dbar(x)-\ubar(x)] < x[\ubar(x)-\sbar(x)]$
since the mass difference between the $u$ and $d$ quarks is far smaller than
that between the $u$ and $s$ quarks.
However, our calculations (see Figs.~2 and 3) show that
$x[\dbar(x)-\sbar(x)] > x[\dbar(x)-\ubar(x)] > x[\ubar(x)-\sbar(x)]$.
The relation $x[\dbar(x)-\ubar(x)] > x[\ubar(x)-\sbar(x)]$
is opposite to our above argument,
which implies that the dynamics responsible for the $SU(3)$ symmetry breaking
in the quark distributions of the $\Sigma^+$ sea, as calculated in our model,
are different from that responsible for the mass differences among the
$u$, $d$ and $s$ quarks.

Another interesting question concerning the $\Sigma^+$ sea is the
quark-antiquark asymmetry.
Although the perturbative sea created from gluon-splitting is symmetric
$q=\qbar$ (in the leading twist approximation in perturbative calculation),
the non-perturbative sea, which may exist over a long time
and has a strong connection with the ``bare" $\Sigma^+$,
may be asymmetric $q\neq \qbar$.
Because of the existence of valance $u$ and $s$ quarks in the $\Sigma^+$,
it is difficult to measure the differences $u-\ubar$ and $s-\sbar$
in the $\Sigma^+$ sea.
The most likely experiment is to measure the difference
between $d$ and $\dbar$.
From the baryon-meson fluctuation model the $d(x)-\dbar(x)$
turns out to be:
\bea
d(x)-\dbar(x)&=& d_{\Lambda \pi^+}(x)-\dbar_{\Lambda \pi^+}(x)
			     +d_{\Sigma^0 \pi^+}(x)-\dbar_{\Sigma^0 \pi^+}(x)
			     +d_{p {\bar K^0}}(x)-\dbar_{p {\bar K^0}}(x) \nonumber \\
&=& \int^1_x \frac{dy}{y}
\left\{
\left[
 f_{\Lambda \pi^+}(y) + f_{\Sigma^0 \pi^+}(y) + f_{p {\bar K^0}}(1-y)
 \right]
 d^p(\frac{x}{y}) 
 \right. \nonumber\\
& & ~~~~~~~
- \left[
 f_{\Lambda \pi^+}(1-y) + f_{\Sigma^0 \pi^+}(1-y) \right]
 \dbar^{\pi^+}(\frac{x}{y}) \nonumber \\
& & ~~~~~~~
\left.
 - f_{p {\bar K^0}}(1-y) \dbar^{\bar K^0}(\frac{x}{y}) \right\}.
\label{dmd}
\eea
The numerical results at scales $\mu^2_{\rm NLO}$ and $Q^2=4$ GeV$^2$
are presented in Fig.~4.
Once again one can find that evolution ``pushes" the
distributions to the small $x$ region.
It can be seen that $d\neq\dbar$ in the $\Sigma^+$ sea.
However, the prediction for the behavior of $d(x)-\dbar(x)$
depends strongly on the value of $\alpha$
-- for $\alpha=0.3$~GeV $d(x) <\dbar(x)$ in the smaller $x$ region and
$d(x) >\dbar(x)$ in the larger $x$ region,
while for the $\alpha=1.0$~GeV $d(x) >\dbar(x)$ in the smaller $x$ region and
$d(x) < \dbar(x)$ in the larger $x$ region.
This result is similar to our earlier finding on the $s(x)-\sbar(x)$
in the nucleon sea \cite{Cao}
employing the same light-cone baryon-meson fluctuation model,
which suggests that $SU(3)$ symmetry in the sea holds in this case.

We turn to the discussion about $\alpha$-dependence in our calculation.
Comparing Figs.~2 and 3 one can find that
for the $x[\dbar(x)-\ubar(x)]$, $x[\dbar(x)-\sbar(x)]$ and $x[\ubar(x)-\sbar(x)]$
the shape and maximum of asymmetries are very similar
for different $\alpha$, while the $x$ position at which the asymmetries
exhibit maxima shifts slightly.
The calculations with $\alpha=0.3$~GeV peak at about $x\simeq 0.06$
while the calculations with $\alpha=1.0$~GeV peak at about $x\simeq 0.1$.
Thus the calculations for the flavour asymmetry are not very sensitive
to the value of $\alpha$, and
$x$ being about $0.08$ is a good region to study the flavor asymmetry
in the $\Sigma^+$ sea.
This observation is consistent with the prediction given in Ref.~\cite{Alberg}
-- the region $0.1\leq x \leq 0.2$ should be a good one to
measure the flavour asymmetry in the $\Sigma$ sea.
The calculation for the $d(x)-\dbar(x)$ (see Fig.~4) is much more
sensitive to the value of $\alpha$ than that for the flavour asymmetry
-- the calculations with $\alpha=0.3$ GeV and $1.0$ GeV even give
opposite predictions for the $x$-dependence of $d(x)-\dbar(x)$.
We may expect that further calculation \cite{CaoPreparation} on the nucleon sea
employing the light-cone baryon-meson fluctuation model
may provide useful constraints on the value of $\alpha$,
and thereby give more definite predictions on
the sea quark content in the $\Sigma^+$ baryon.

In summary, besides the nucleon sea
the studies on the sea quark content of the other members of the baryon octet
are interesting and important since it is helpful to our understanding of
both the structure of the octet baryons and non-perturbative QCD effects
such as $SU(3)$ symmetry breaking and flavour asymmetry.
We calculated the sea quark content of the $\Sigma^+$ baryon employing
the light-cone baryon-meson fluctuation model.
It was found that the $\Sigma^+$ sea is flavour asymmetric
($\dbar > \ubar> \sbar$) and quark-antiquark asymmetric
($q \not= \qbar$).
Our prediction for the flavor asymmetry, $\dbar > \ubar > \sbar$, is significantly
different from the $SU(3)$ prediction ($\dbar < \ubar < \sbar$),
while our prediction for the $d$-$\dbar$ asymmetry is consistent with the
$SU(3)$ prediction.

\section*{Acknowledgments}
This work was partially supported by the Massey Postdoctoral Foundation, New Zealand.

\newpage
\begin{center}
{ Table 1. Parameters in Eq.~(\ref{qfit})} at $Q^2=4$ GeV$^2$.

\vskip 0.5cm
\begin{tabular}{|c|c|c|c|c|c|}\hline
   & $a$ & $b$ & $c$ & $d$ & $e$ \\ \hline
$v^\pi(x,Q^2)$ & $1.712$ & $-0.518$ & $1.182$ & $-0.836$ & $0.972$ \\ \hline
$\dbar^{{\bar K^0}}(x, Q^2)$ & $0.910$ & $-0.519$ & $1.418$ & $-0.910$ & $1.086$
 \\ \hline 
$d^p(x, Q^2)$ & $0.615$ & $-0.575$ & $5.096$ & $1.102$ & $6.773$ \\ \hline
\end{tabular}
\end{center}

\newpage
\section*{Figure Captions}
\begin{description}
\item
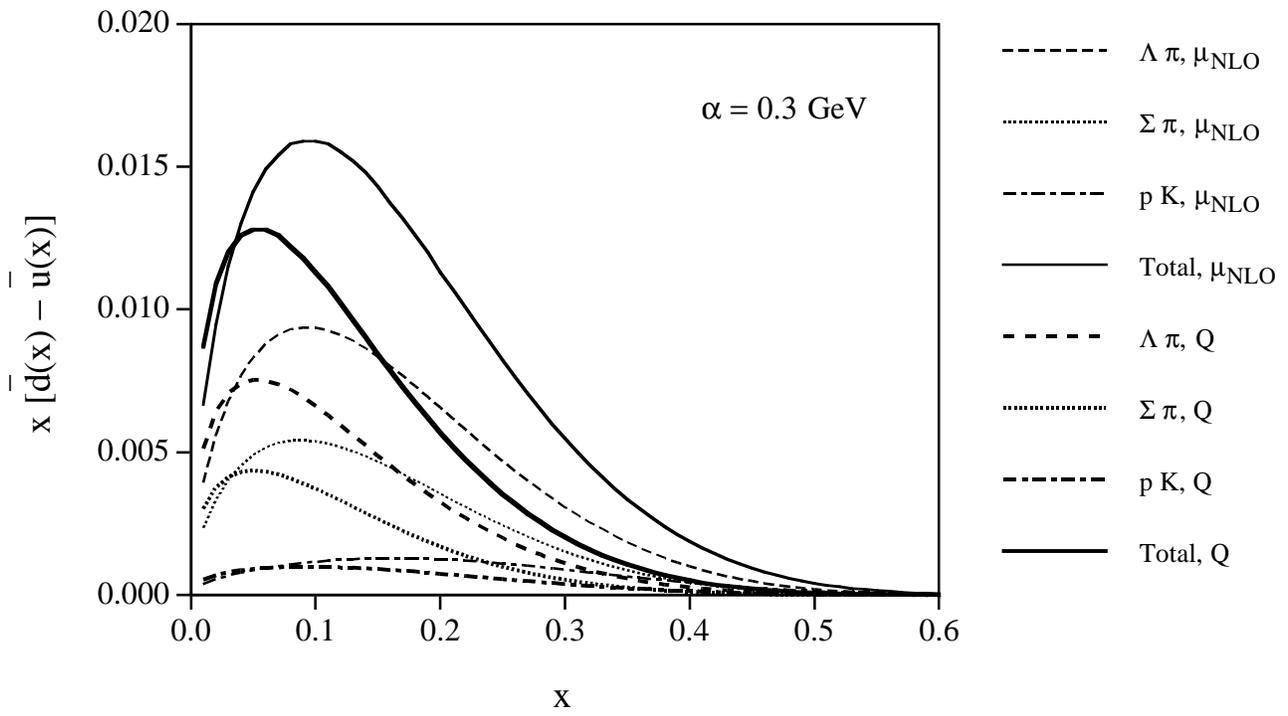
{Fig.~1.} 
$x[\dbar(x)-\ubar(x)]$ with $\alpha=0.3$ GeV.
The dashed, dotted, and dashed-dotted curves are
the contributions from $\Lambda \pi$,
$\Sigma \pi$ and $p K$ components respectively.
The solid curve is the sum of above three contributions.
The thinner and thicker curves correspond to the scales
$\mu^2_{\rm NLO}=0.34$ GeV$^2$ and $Q^2=4$ GeV$^2$ respectively.
\item
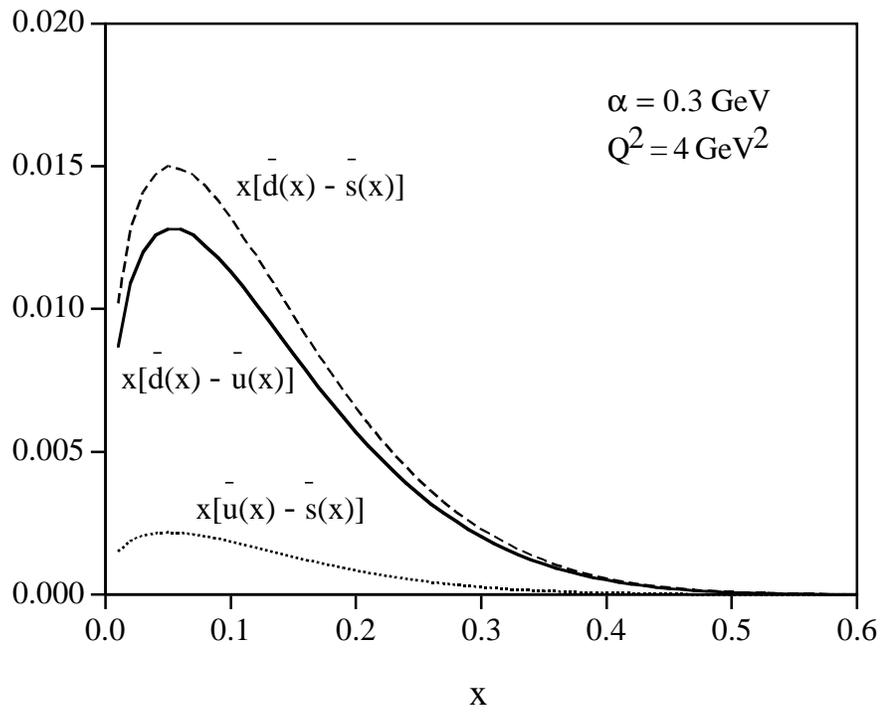
{Fig.~2.}
$x[\dbar(x)-\ubar(x)]$, $x[\dbar(x)-\sbar(x)]$
and $x[\ubar(x)-\sbar(x)]$ at $Q^2=4$ GeV$^4$
and with $\alpha=0.3$ GeV.
\item
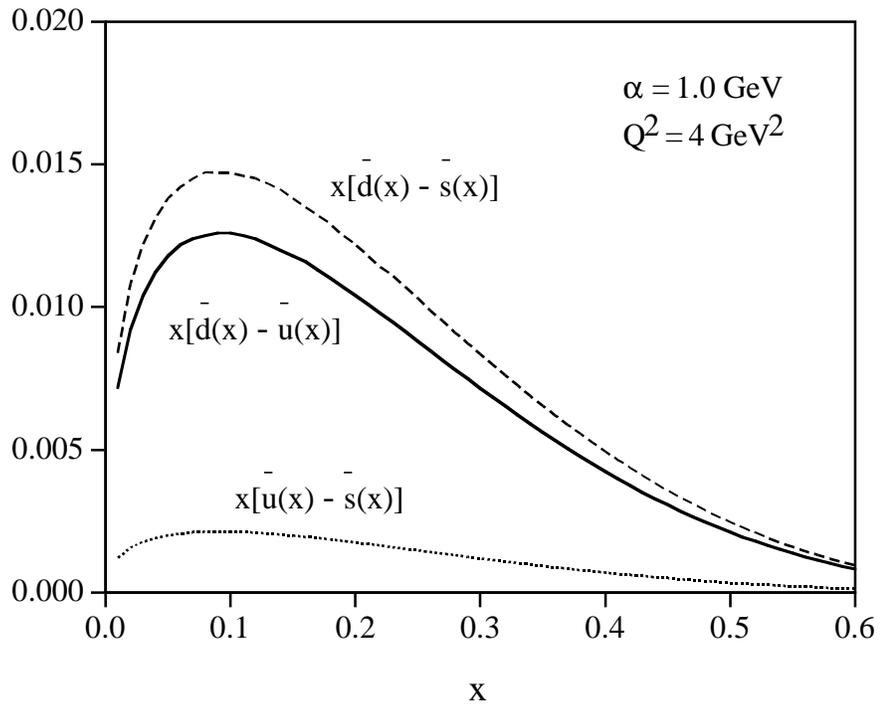
{Fig.~3.}
Same as Fig.~2 but with
$\alpha=1.0$ GeV.
\item
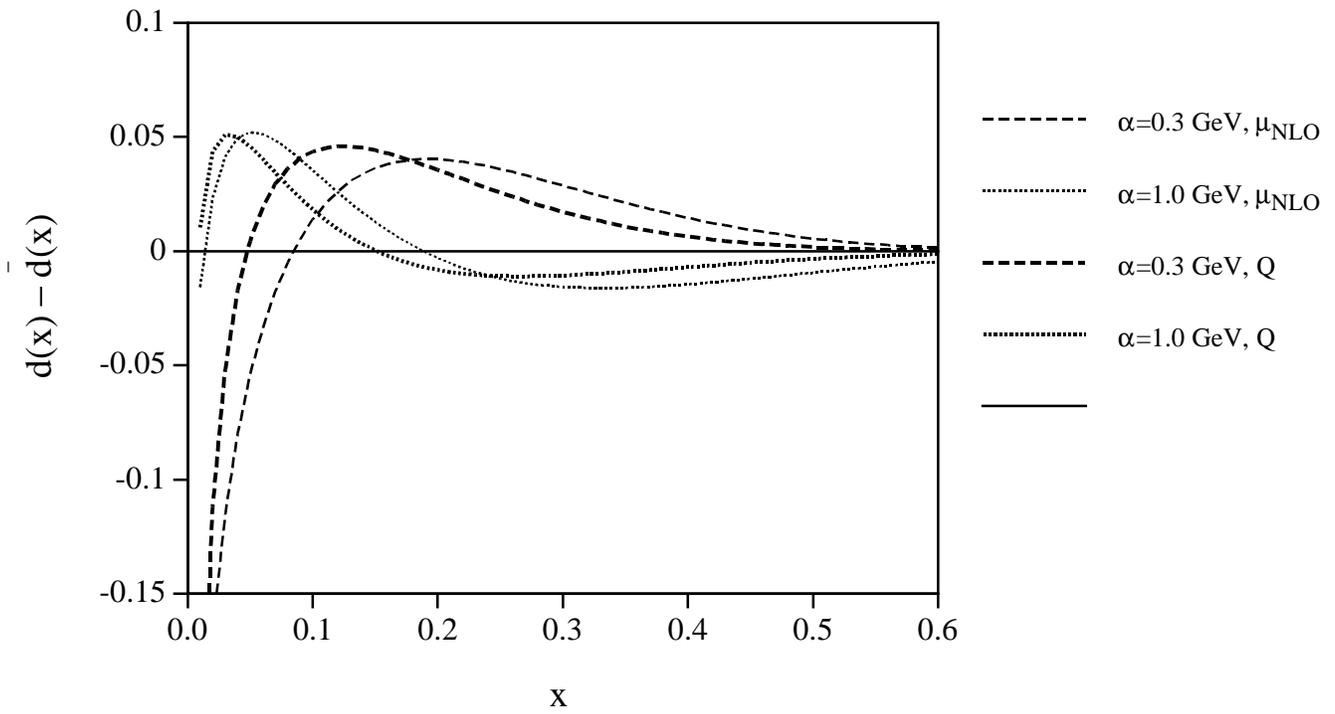
{Fig.~4}
$d(x)-\dbar(x)$ with $\alpha=0.3$ GeV (dashed curve)
and $\alpha=1.0$ GeV (dotted curve).
The thinner and thicker curves correspond to the scales
$\mu^2_{\rm NLO}$ and $Q^2=4$ GeV$^2$ respectively.
\end{description}

\end{document}